\documentclass[fleqn,usenatbib,nofootinbib]{mnras}
\usepackage{graphicx}
\usepackage{amsmath,amssymb}
\usepackage{lastpage}
\usepackage[all]{hypcap}
\usepackage[T1]{fontenc}
\usepackage{float}
\usepackage{subfigure}
\usepackage{afterpage}
\usepackage[usenames]{color}
\usepackage{mathrsfs}
\usepackage{upgreek}
\usepackage{hyperref}
\usepackage{gensymb}
\usepackage{color}
\usepackage[dvipsnames]{xcolor}
\usepackage{longtable}
\usepackage{multicol}
\usepackage{threeparttable}
\usepackage{multirow}
\usepackage{placeins}
\usepackage{bm}
\usepackage[capitalise]{cleveref}
\usepackage{enumitem}
\usepackage{mathtools}
\usepackage{booktabs}

\graphicspath{{Figures/}}

\usepackage{fontawesome5}
\newcommand{\github}[1]{\href{#1}{\faGithubSquare}}
\newcommand{\githublink}{\github{https://github.com/harrydesmond/GalaxySpinAnisotropy}}

\newcommand{\icg}{Institute of Cosmology \& Gravitation, University of Portsmouth, Dennis Sciama Building, Portsmouth, PO1 3FX, UK}

\title[Galaxies spin isotropically]{No evidence for anisotropy in galaxy spin directions}
\author[Patel \& Desmond]{
Dhruva~Patel$^{1}$\thanks{zx970439@ou.ac.uk} and Harry~Desmond$^{1}$\thanks{harry.desmond@port.ac.uk}\\
$^{1}$\icg\\
}

\pubyear{2024}

\begin{document}
\label{FirstPage}
\pagerange{\pageref{FirstPage}--\pageref{LastPage}}
\maketitle

\begin{abstract}
Modern cosmology rests on the \emph{cosmological principle}, that on large enough scales the Universe is both homogeneous and isotropic. A corollary is that galaxies' spin vectors should be isotropically distributed on the sky. This has been challenged by multiple authors for over a decade, with claims to have detected a statistically significant dipole pattern of spins. We collect all publicly available datasets with spin classifications (binary Z-wise/S-wise), and analyse them for large-angle anisotropies ($\ell \le 2$). We perform each inference in both a Bayesian and frequentist fashion, the former establishing posterior probabilities on the multipole parameters and the latter calculating $p$-values for rejection of the null hypothesis of isotropy (i.e. no power at $\ell>0$). All analysis indicate consistency with isotropy to within $3\sigma$. We similarly identify no evidence for a ``hemisphere anisotropy'' that neglects the angular dependence of the dipole. We isolate the differences with contrary claims in the ad hoc or biased statistics that they employ. Our code is publicly available \githublink.
\end{abstract}

\begin{keywords}
galaxies: formation -- galaxies: fundamental parameters -- galaxies: statistics -- large-scale structure of Universe
\end{keywords}

\section{Introduction}
\label{sec:intro}

When averaged over sufficiently large scales, the Universe is believed to be described by General Relativity and the Friedmann--Robertson--Walker metric in which all regions of space and all lines of sight from any observer are equivalent. The homogeneity scale may already be reached at $\sim$70 Mpc, in agreement with concordance $\Lambda$ Cold Dark Matter ($\Lambda$CDM) cosmology \citep{homo_1,homo_4,homo_2,homo_3}. The observational evidence for isotropy is somewhat weaker, and in fact several observations suggest that preferred directions do exist in the Universe. These include anomalies in the Cosmic Microwave Background (CMB; most recently~\citealt{CMB_anisotropy}, although see~\citealt{Sravan}), non-negligible multipoles in the large-scale velocity field traced by supernovae~\citep{Basheer,Hu}, strong bulk flows extending to 100s of Mpc \citep{CF4} and non-convergence of the rest frames of the CMB and distant matter \citep{Rameez,Migkas,EB_1,EB_2,SN_1,SN_2}. For a review of the observational status of the cosmological principle, see~\citet{CP_review}. We must assess carefully whether the fundamental tenets of $\Lambda$CDM hold before we can settle into an era of ``precision cosmology''.

We investigate here a subset of the claims for anisotropy, namely the putative presence of a dipole in galaxies' spin directions when viewed from the Milky Way. This is a clean test with few possible systematics: one uses images of low-inclination late-type galaxies to determine (e.g. from the direction of spiral arm winding) whether they are spinning towards or away from us, and then ask whether this binary-valued field projected onto the sky has significant power in multipoles beyond $\ell=0$ (the monopole). In galaxies, we designate clockwise rotation as Z-wise and counterclockwise rotation as S-wise. A Z-wise galaxy rotates clockwise, with its angular momentum pointing away from us, while an S-wise galaxy rotates counterclockwise, with its angular momentum vector pointing towards us. Provided the galaxies are at cosmological distance, power should not be generated at low $\ell$ from tidal torque-like interactions~\citep{TTT}. If true, this finding would therefore force a rethink of basic cosmology, and may imply that the Universe possessed a net angular momentum in its initial conditions (e.g.~\citealt{ani_1,ani_2,ani_3}).

Over the past $\sim$15 years (although see~\citealt{Dodd,Iye_Sugai,Sugai_Iye} for earlier related attempts) this test has been performed with various datasets, methods for determining spin direction and statistics for quantifying the anisotropy. Although dominated by a few authors, most studies claim to find a significant dipole
(\citealt{Dodd, 2007astro.ph..3694L, Longo, 2017PASA...34...44S, 2020OAst...29...15S, 2020PASA...37...53S, 2020AN....341..324S,2020Ap&SS.365..136S, Shamir_2021, 2021PASA...38...37S, 2022JApA...43...24S, 2022MNRAS.516.2281S, 2022AdAst2022E...5S, McAdam_Shamir_2023_GAN, 2024arXiv240317271S}). On the other hand, \citet{Iye_Sugai}, \citet{Land}, \citet{2017MNRAS.466.3928H}, \citet{Tadaki} and \citet{Iye2021} do not.
Hence this important issue remains controversial.

We collect all publicly available catalogues for which galaxy spin directions have been estimated, a procedure called ``annotation''. We assume these are correct, and question merely the statistics with which this data is interrogated for anisotropy. If we find a dipole we may wonder whether the annotation method suffers from a systematic that causes this, but if we do not find a dipole it is highly unlikely that an existing dipole is hidden by such a systematic. Unlike almost all previous authors we do not use $\chi^2$ to avoid assumptions of Gaussianity.
Instead we define a likelihood for each galaxy's spin as a function of low-$\ell$ multipole parameters (monopole, dipole, quadrupole) and the angle between the galaxy direction and multipole axes. We derive posterior probability distributions on these parameters in a Bayesian analysis and use mock data generated under the assumption of isotropy to test that null hypothesis in a frequentist fashion.

In Sec.~\ref{sec:data} we describe the annotated galaxy catalogues that we employ. In Sec.~\ref{sec:method} we detail our methods, separately for the Bayesian and frequentist approach. Sec.~\ref{sec:results} presents our results and Sec.~\ref{sec:conc} concludes.

\section{Observational Data}
\label{sec:data}

We collate all publicly available image data that has been used in the literature to test the isotropy of galaxy spins. Beyond the raw data, this test requires an algorithm to calculate the spin direction of each galaxy (annotation). Any difference in results for fixed data could arise either from the annotation method or the statistics with which the annotated data is tested for anisotropy. Here we accept at face value the annotation of the utilised datasets by other authors, and ask merely whether the statistics of the annotated datasets provide compelling evidence for anisotropy. While the annotations themselves may of course be biased (we refer to the relevant papers for arguments that they are unlikely to be), if they imply isotropy it seems highly unlikely that such biases would hide an underlying anisotropy, which one would expect them if anything to increase.

\begin{figure*}
  \centering
  \includegraphics[width=0.99\textwidth]{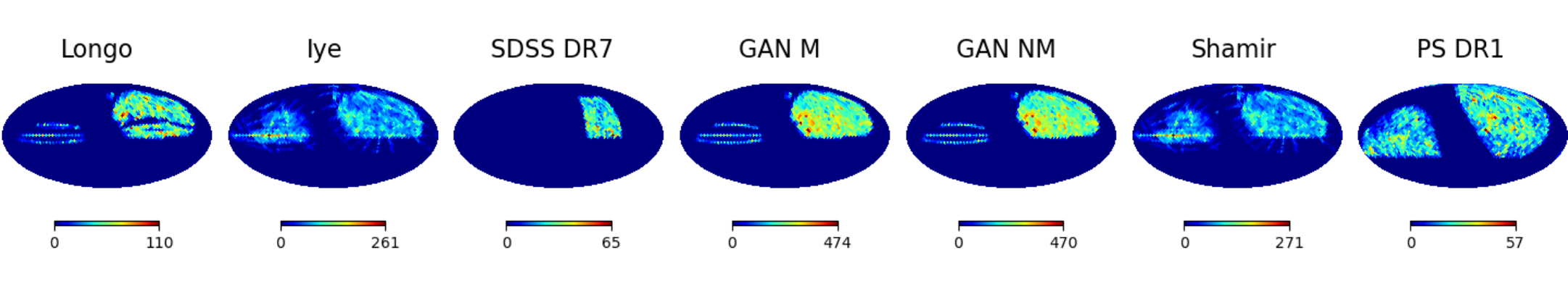}
  \caption{The number of galaxies per pixel for each of the datasets we investigate. These are Mollweide projections in equatorial coordinates using \textsc{healpix} with \texttt{nside} $=16$, rotated to centre on (RA, Dec)$=(\frac{\pi}{2}, 0)$ and RA increases towards the left.}
  \label{fig:skyplot}
\end{figure*}

The datasets we use are summarised in Table \ref{Table 1}. Most of the datasets (\texttt{Longo}, \texttt{Iye}, \texttt{SDSS DR7}, \texttt{GAN M}, \texttt{GAN NM} and \texttt{Shamir}) come from data releases 6-8 of the Sloan Digital Sky Survey (SDSS; \citealt{SDSS, SDSS_DR8}).
\texttt{PS DR1} derives instead from the Panoramic Survey Telescope and Rapid Response System (Pan-STARRS1) data release 1 \citep{PS1}. This was obtained by cross-matching galaxies with identical IDs between two Pan-STARRS datasets, \citet{2017PASA...34...44S} and \citet{2020ApJS..251...28G}. Thus, this dataset is to our knowledge unique, and no isotropy analysis has previously been conducted on it (although it was annotated in \citealt{2017PASA...34...44S}).
The SDSS datasets differ in sky coverage and galaxy density as seen in Figure \ref{fig:skyplot}. \texttt{GAN M} is almost identical to \texttt{GAN NM} except the galaxy images were mirrored before being fed into the annotation algorithm, in order to quantify the level of asymmetry in this algorithm. The quoted sigma values in Table \ref{Table 1} were taken from the cited papers, except in the case of \citep{Longo} where it was calculated from the $p$-value quoted in the abstract of that paper assuming a Gaussian distribution and the sigma value given for \texttt{Iye} was taken from \citep{2022PASJ...74.1114S} as it is the more significant result in the analysis of this dataset.

Various annotation methods were used. \cite{Longo} employed a group of undergraduate students, referred to as ``scanners'', to manually annotate randomly assigned redshift  slices of the data. The author states that any proclivity for the scanners to prefer a particular spin direction was mitigated by mirroring half of the objects at random to disfavour a particular handedness. The remaining datasets were annotated either by \textsc{SpArcFiRe} (``Scalable Automated Detection of Spiral Galaxy Arm Segments; \citealt{2014ApJ...790...87D}), an algorithm which extracts the structural features of spiral galaxies, or \textsc{Ganalyzer} (``Galaxy Analyzer''), a modelling tool for automated galaxy classification
\citep{2011ApJ...736..141S}. We investigate the consistency of different annotation methods by cross-matching galaxies with identical
IDs between the SDSS-based datasets, finding agreement in spin direction for 91.81\% of galaxies matched between \texttt{Longo} and \texttt{GAN M}. As the latter dataset is fully mirrored, this implies that the former is also. This is corroborated by an 8.27\% agreement between \texttt{Longo} and \texttt{GAN NM}, and a 93.36\% agreement between \texttt{GAN NM} and \texttt{SDSS DR7}. The level of mirroring is however not important for our analysis, which aims simply to investigate the statistical significance for anisotropy from a given set of spin values.

We visualise the datasets in Fig.~\ref{fig:skyplot} by plotting the number of galaxies per pixel under a \textsc{healpix} scheme with \texttt{nside}$=16$. We see a significant overlap in area between most of the datasets in the SDSS region. It is clearly imperative for the statistical method used to assess anisotropy to be robust to a highly incomplete sky coverage.

\begin{table}
\centering
\caption{The observational datasets we use to search for galaxy spin anisotropy. The third column gives the significance of anisotropy reported by the creators of each dataset, where such an analysis was performed. M denotes (partial) mirroring of the data, while the final column gives the annotation algorithm.
In order, the references are \citet{Longo, 2021ApJ...907..123I, 2022PASJ...74.1114S, McAdam_Shamir_2023_SDSS_DR7, McAdam_Shamir_2023_GAN, Shamir_2021, 2017PASA...34...44S}. Note, \texttt{GAN M} and \texttt{GAN NM} are cited from the same paper \citep{McAdam_Shamir_2023_GAN}.}
\begin{tabular}{|c|c|c|c|c|}
\hline
\textbf{Name} & \textbf{\# gals} & $\mathbf{\sigma}$ & $\mathbf{M}$ & \textbf{Annotation method} \\
\hline
\texttt{Longo} & 15158 & 3.16 & Yes & Human scanners \\
\texttt{Iye} & 72888 & 2.10 & No & \textsc{Ganalyzer} \\
\texttt{SDSS DR7} & 6103 & --- & No & \textsc{Ganalyzer} \\
\texttt{GAN M} & 139852 & 3.97 & Yes & \textsc{SpArcFiRe}, \textsc{Ganalyzer} \\
\texttt{GAN NM} & 138940 & 2.33 & No & \textsc{SpArcFiRe}, \textsc{Ganalyzer} \\
\texttt{Shamir} & 77840 & 2.56 & No & \textsc{Ganalyzer} \\
\texttt{PS DR1} & 28731 & --- & No & \textsc{Ganalyzer} \\
\hline
\end{tabular}
\label{Table 1}
\end{table}

\section{Method}
\label{sec:method}

To ensure that our results are robust to choice of methodology---and suit the taste of the reader---we perform both a Bayesian and frequentist analysis. Each of these relies on a function that describes the likelihood of the data given the model parameters. These parameters, which we denote $\vec{\theta}$, are (some subset of) monopole magnitude $M$, dipole magnitude $D$ and unit vector direction on the sky $\vec{d} = \{d_\alpha, d_\delta\}$, and quadrupole magnitude $Q$ with corresponding unit sky vectors $\vec{q}_1 = \{q_{1,\alpha}, q_{1,\delta}\}$ and $\vec{q}_2 = \{q_{2,\alpha}, q_{2,\delta}\}$. These are multipoles of the on-sky probability field for spins to be Z-wise as seen from the Milky Way: a monopole describes a direction-independent preference for a particular spin direction (most likely due to a bias in the annotation method), a dipole describes a preference for Z-wise spins in one direction and S-wise in the opposite, and a quadrupole describes a pattern with two hotspots and corresponding coldspots. We work in equatorial coordinates, where $\alpha$ denotes right ascension (RA) and $\delta$ declination (Dec). We adopt the range of $[0, 2\pi)$ for RA and $(-\frac{\pi}{2}, \frac{\pi}{2})$ for Dec throughout. We denote a galaxy's spin value as $s$, which we assign the value 0 if the spin is S-wise as seen from the Milky Way, and 1 if it is Z-wise. Isotropy therefore corresponds to $D=Q=0$, and an equal number of Z-wise and S-wise spins to $M=0.5$.

For galaxy $i$, the likelihood of Z-wise spin is
\begin{equation}\label{eq:L}
\mathcal{L}(s_i|\vec{\theta}) = M + D \: \vec{d} \cdot \vec{n}_i + Q \left(\vec{q}_1 \cdot \vec{n}_i \: \vec{q}_2 \cdot \vec{n}_i - \frac{1}{3} \vec{q}_1 \cdot \vec{q}_2\right),
\end{equation}
where $\vec{n}$ is the unit vector pointing in the direction of the galaxy. The likelihood of S-wise spin is $1-\mathcal{L}$, and thus it is properly normalised by construction provided $0 \le \mathcal{L} \le 1$. We impose this by requiring $0 \le |M|+|D|+|Q| \le 1$, but find that this never comes into play because the best-fit $M$ values are $\sim$0.5 while the best-fit $D$ and $Q$ values are small.
This matches the model of~\citet{Land}.
We assume that all galaxies in a dataset are independent, so that the likelihood of the dataset is the product of the likelihood of its constituent galaxies.
To investigate how the results are affected by the inclusion of the $\ell=0$, $\ell=1$ and $\ell=2$ terms we perform separate analyses modelling i) monopole only, ii) dipole only at $M=0.5$, iii) monopole and dipole, and iv) monopole, dipole and quadrupole.

\begin{table}\label{tab:params}
\centering
\caption{Table of free parameters, descriptions and their prior ranges. $[\cos(d_\delta)]^{+\pi/2}_{-\pi/2}$ denotes a prior proportional to $\cos(d_\delta)$ within the range $-\pi/2 \le d_\delta \le \pi/2$. Below the horizontal line we show the parameters (besides $M$) of the alternative hemisphere anisotropy analysis (see final paragraph of Sec.~\ref{sec:meth_bayes}), which replace $D$, $d_\alpha$ and $d_\delta.$}
\label{tab:M}
\renewcommand{\arraystretch}{1.5} 
\begin{tabular}{|l|c|c|}
\hline
\textbf{Parameter} & Description & Prior \\
\hline
\texttt{$M$} & Monopole magnitude & $\mathcal{U}(0.3, 0.7)$ \\
\texttt{$D$} & Dipole magnitude & $\mathcal{U}(0,0.3)$ \\
\texttt{$Q$} & Quadrupole magnitude & $\mathcal{U}(0,0.3)$ \\
\texttt{$d_{\alpha}$} & Unit dipole RA component & $\mathcal{U}(0,2\pi)$ \\
\texttt{$d_{\delta}$} & Unit dipole Dec component & $[\cos(d_\delta)]^{+\pi/2}_{-\pi/2}$ \\
\texttt{$q_{1,\alpha}$} & Unit quadrupole $1^\text{st}$ axis RA & $\mathcal{U}(0,2\pi)$ \\
\texttt{$q_{1,\delta}$} & Unit quadrupole $1^\text{st}$ axis Dec & $[\cos(q_{1,\delta)}]^{+\pi/2}_{-\pi/2}$ \\
\texttt{$q_{2,\alpha}$} & Unit quadrupole $2^\text{nd}$ axis RA & $\mathcal{U}(0,2\pi)$ \\
\texttt{$q_{2,\delta}$} & Unit quadrupole $2^\text{nd}$ axis Dec & $[\cos(q_{2,\delta)}]^{+\pi/2}_{-\pi/2}$ \\
\hline
\texttt{$A$} & Asymmetry magnitude & $\mathcal{U}(0,0.3)$ \\
\texttt{$a_{\alpha}$} & Unit asymmetry RA component & $\mathcal{U}(0,2\pi)$ \\
\texttt{$a_{\delta}$} & Unit asymmetry Dec component & $[\cos(a_\delta)]^{+\pi/2}_{-\pi/2}$ \\
\hline
\end{tabular}
\end{table}

\subsection{Bayesian analysis}
\label{sec:meth_bayes}

The goal of a Bayesian analysis is to establish posterior probabilities on the model parameters. We adopt uniform priors on $M$, $D\ge0$ and $Q\ge0$,
and a uniform prior on area element for the $\vec{d}$, $\vec{q}_1$ and $\vec{q}_2$ vectors across the full sky. This corresponds to a prior uniform in the vector's RA components
and in the sine of their Dec components.
See Table 2 for a description of the free parameters used in the analysis, and their priors for the Bayesian analysis.
We initially choose the prior ranges in Table 2, with the intention of expanding the ranges if necessary, but find that the posteriors are already contained within these bounds in all cases.
To expedite sampling and eliminate multimodality, we break the symmetry between the two quadrupole vectors by requiring $q_{2,\alpha} \ge q_{1,\alpha}$.

We perform a Markov Chain Monte Carlo (MCMC) analysis with the affine-invariant sampler \texttt{emcee} \citep{emcee}, using 22 walkers with initial positions randomly sampled from the prior. We calculate the autocorrelation length for each parameter every 100 iterations, terminating when the chain is at least 100 autocorrelation lengths in each parameter and the change in autocorrelation length between iterations is less than 1 per cent.

This produces corner plots describing the posteriors on the parameters and their degeneracies. We summarise each marginal posterior using its mode $\bar{\theta}$ and 68 per cent confidence interval,
unless $\bar{\theta}-2\:\text{std}(\theta)<0$ in which case we instead quote only the 68 per cent upper limit. The mode is calculated by \texttt{emcee} as the location of the bin containing the most samples. The use of a 68$\%$ confidence interval corresponds to 1$\sigma$ for a Gaussian distribution. For any parameter required to be $\geq 0$ by the prior, if it is consistent with 0 (estimated by  $\bar{\theta}-2\:\text{std}(\theta)<0$) we quote an upper limit, while if it is not we quote the mode of the distribution and 68\% confidence interval. This approach ensures that the reported values appropriately reflect the characteristics of the posterior distribution and the constraints imposed by the prior. We assess the goodness-of-fit of each model using the Bayesian information criterion (BIC) as an approximation to the Bayesian evidence. This is given by~\citet{BIC}:
\begin{equation}
\text{BIC} \equiv k\ln(N) - 2\ln(\hat{\mathcal{L}})
\end{equation}
where $k$ is the number of free parameters, $N$ the number of data points and $\hat{\mathcal{L}}$ the maximum-likelihood value. The BIC shows whether the addition of parameters is warranted by the data: an extra parameter must increase the maximum likelihood by at least $\ln(N)/2$. As the absolute value is unimportant, we show only differences ($\Delta$BIC) relative to the baseline model inferring $M$ only.

Finally, we run a separate analysis with a model that investigates the preference for observing a particular spin direction in a given hemisphere, without any variation in predicted spin within the hemisphere. We dub this a ``hemisphere anisotropy'' to distinguish it from a dipole.
In this model, a galaxy with unit direction $\vec{n}$ on the sky has a likelihood of Z-wise spin given by
\begin{equation}\label{eq:A}
\mathcal{L}(s|\vec{\theta}) =
\begin{cases}
M + A & \text{if } \vec{n} \cdot \vec{a} > 0\\
M - A & \text{otherwise}
\end{cases}
\end{equation}
where $\vec{a}$ is the unit hemisphere axis, $A$ is the strength of the anisotropy and $M$ is the monopole as before.
The likelihood of S-wise spin is $1-\mathcal{L}$, ensuring proper normalisation provided $0 \le \mathcal{L} \le 1$. We impose this by requiring $0 \le |M|+|A| \le 1$.
We adopt uniform priors on ${M}$ and ${A} \geq 0$, and a uniform prior on area element for vector $\vec{a}$ (see Table~\ref{tab:params}). We perform a similar Bayesian analysis for this model, running MCMC and calculating $\Delta$BIC relative to the same baseline model inferring $M$ only (the identical isotropic model to the dipole analysis). This is designed to mimic the type of anisotropy studied in \citet{2024arXiv240317271S}.

\subsection{Frequentist analysis}
\label{sec:meth_freq}

The goal of a frequentist analysis is to calculate a $p$-value for rejection of a null hypothesis, in this case that the Universe is isotropic. First we calculate the maximum-likelihood values of $\vec{\theta}$ for each dataset using the Nelder--Mead algorithm \citep{nelder1965simplex, gao2012implementing}. Then, for each sample of Table~\ref{Table 1}, we create 50,000 mock datasets\footnote{We find the $p$-values to converge quite slowly with number of mock datasets, hence the large number required for near-stability at two significant figures. The conclusion that the anisotropy is insignificant is readily apparent using fewer mock datasets, however.} with galaxies in the same positions as in the real data but the spins randomised. As we are interested in testing isotropy and not a direction-independent preference for Z-wise or S-wise spins (which is what a bias in annotation method would naturally produce), the mock data is generated using the maximum-likelihood $M$ value, $\widehat{M}$, from the monopole plus dipole model, but $D=Q=0$. We refit each mock data set to calculate the maximum-likelihood $\vec{\theta}$, and then calculate the $p$-value of the null hypothesis as the fraction of mock datasets with more extreme $\{M,D\}$ values than the real data. This is done by binning the mock data in the $\{M,D\}$ plane and calculating contour levels minimally enclosing fixed fractions of the mock datasets; the contour passing through the real-data point determines the $p$-value. In this case we do not consider a quadrupole. \citet{Iye2021} utilised mock data by performing 50,000 Monte Carlo simulations, randomly assigning spin directions to each galaxy to create a baseline distribution of dipole amplitude for an isotropic distribution. By comparison, the observed dipole amplitude was calculated by taking the vector sum of spin directions weighted by their positions on the sky and a 3D random walk model was used to represent the isotropic distribution of galaxy spins, with each spin direction considered as a step in the walk.

\subsection{Validation}
\label{sec:validation}

\begin{figure}
  \centering
  \includegraphics[width=0.49\textwidth]{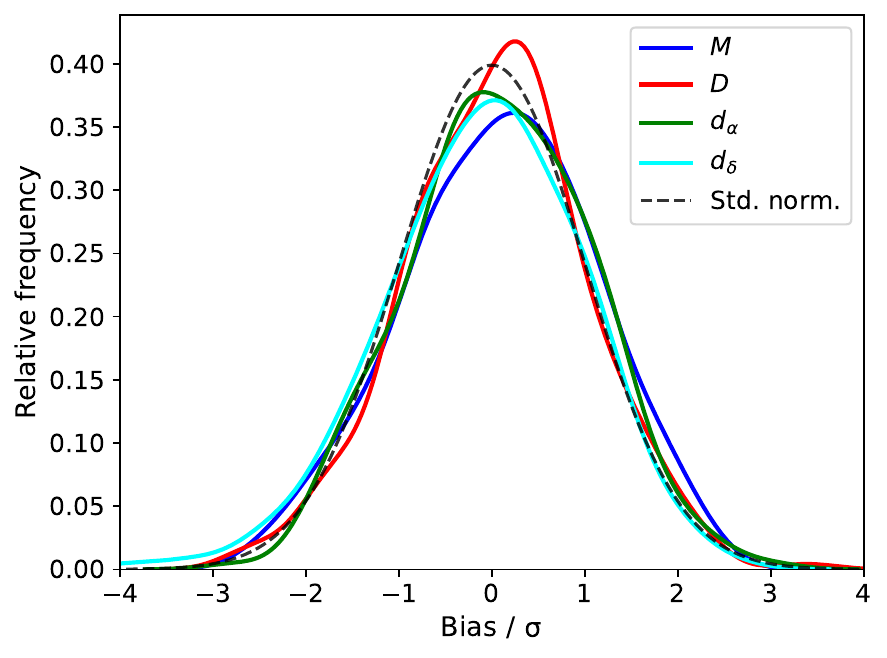}
  \caption{Distribution of bias values (Eq.~\ref{eq:bias}) from MCMC analyses of 300 mock datasets generated by $\tilde{M}=0.6$, $\tilde{D}=0.2$, $\tilde{d}_\alpha=\pi$, $\tilde{d}_\delta=-\pi/4$. The values are expected to follow a standard normal distribution, shown in dashed black.}
  \label{fig:biases}
\end{figure}

Before applying our method to the real data we validate it on mock data to ensure that it returns unbiased parameter values. Each mock dataset has the same number of galaxies as \texttt{Iye} (72888), but we generate mock spin values and optionally randomise the positions of the galaxies on the sky. The mock spin values are generated stochastically according to the probabilities corresponding to some true, generating $\vec{\theta}$. We calculate a bias value for each parameter and each dataset as
\begin{equation}\label{eq:bias}
\text{bias} \equiv \frac{(\langle{\theta}\rangle - \tilde{\theta})}{\text{std}(\theta)},
\end{equation}
following the Bayesian setup, where angular brackets denote the mean and tilde the true, generating value. This may be interpreted as a discrepancy in $\sigma$ between the input parameter value and that recovered by the inference. We find that the distribution of bias values in all cases follows closely the expected standard normal distribution regardless of $\vec{\tilde{\theta}}$ or the positions of the galaxies on the sky. This is illustrated in Fig.~\ref{fig:biases} for the case $\tilde{M}=0.6$, $\tilde{D}=0.2$, $\tilde{d}_\alpha=\pi$, $\tilde{d}_\delta=-\pi/4$ without randomising galaxy positions, over 300 mock datasets.

Note that both of our methods account for the ``look-elsewhere effect'' that comes into play when testing multiple hypotheses (in this case many possible dipole directions). In the frequentist approach this is accounted for by calculating significance with respect to mock data that has the same properties as the real data and has been processed identically, while in the Bayesian approach it is accounted for by the priors, which appropriately weight the probability that an axis should point in any particular direction.

\section{Results}
\label{sec:results}

\subsection{Bayesian analysis}
\label{sec:res_bayes}

\begin{figure*}
  \centering
  \includegraphics[width=0.48\textwidth]{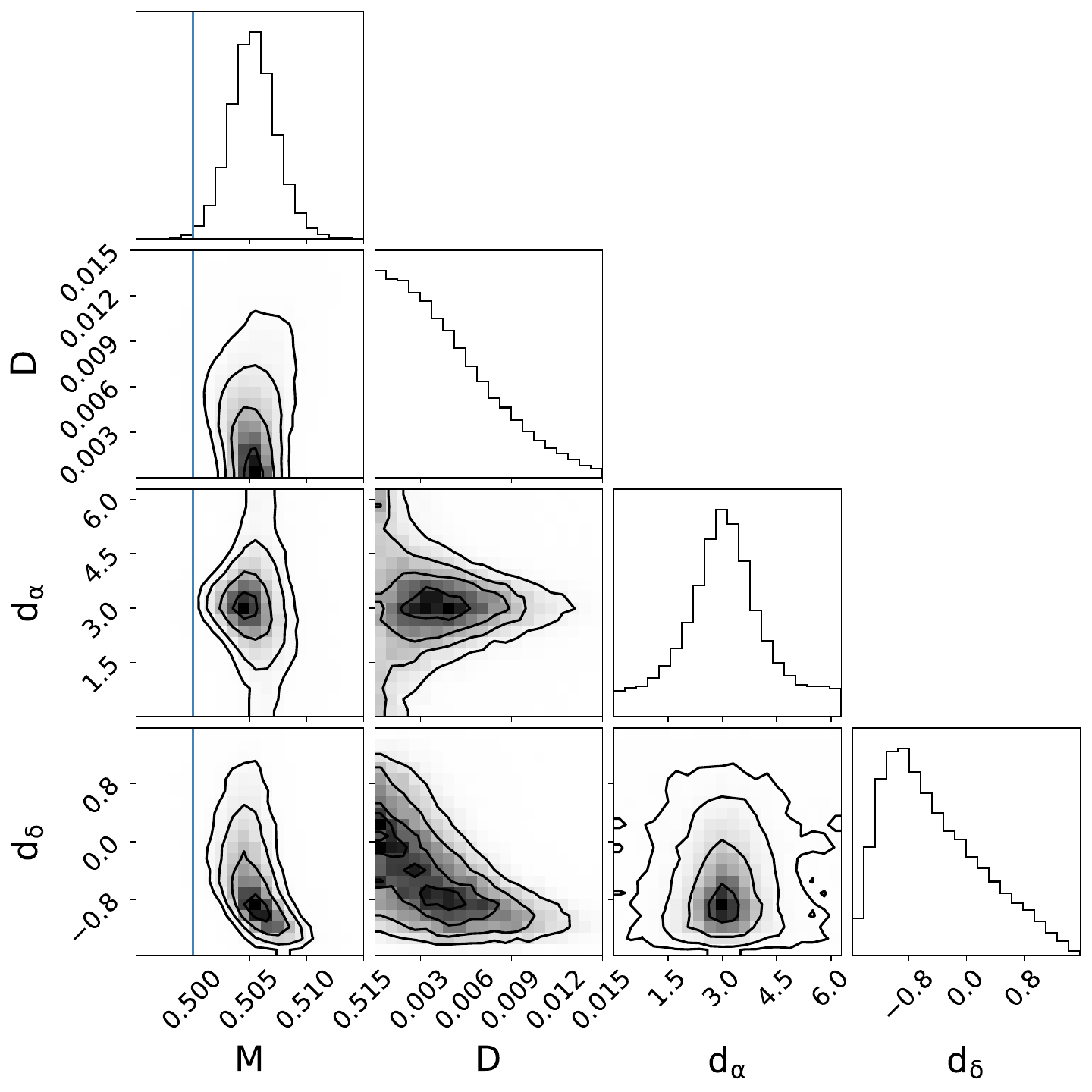}
  \includegraphics[width=0.48\textwidth]{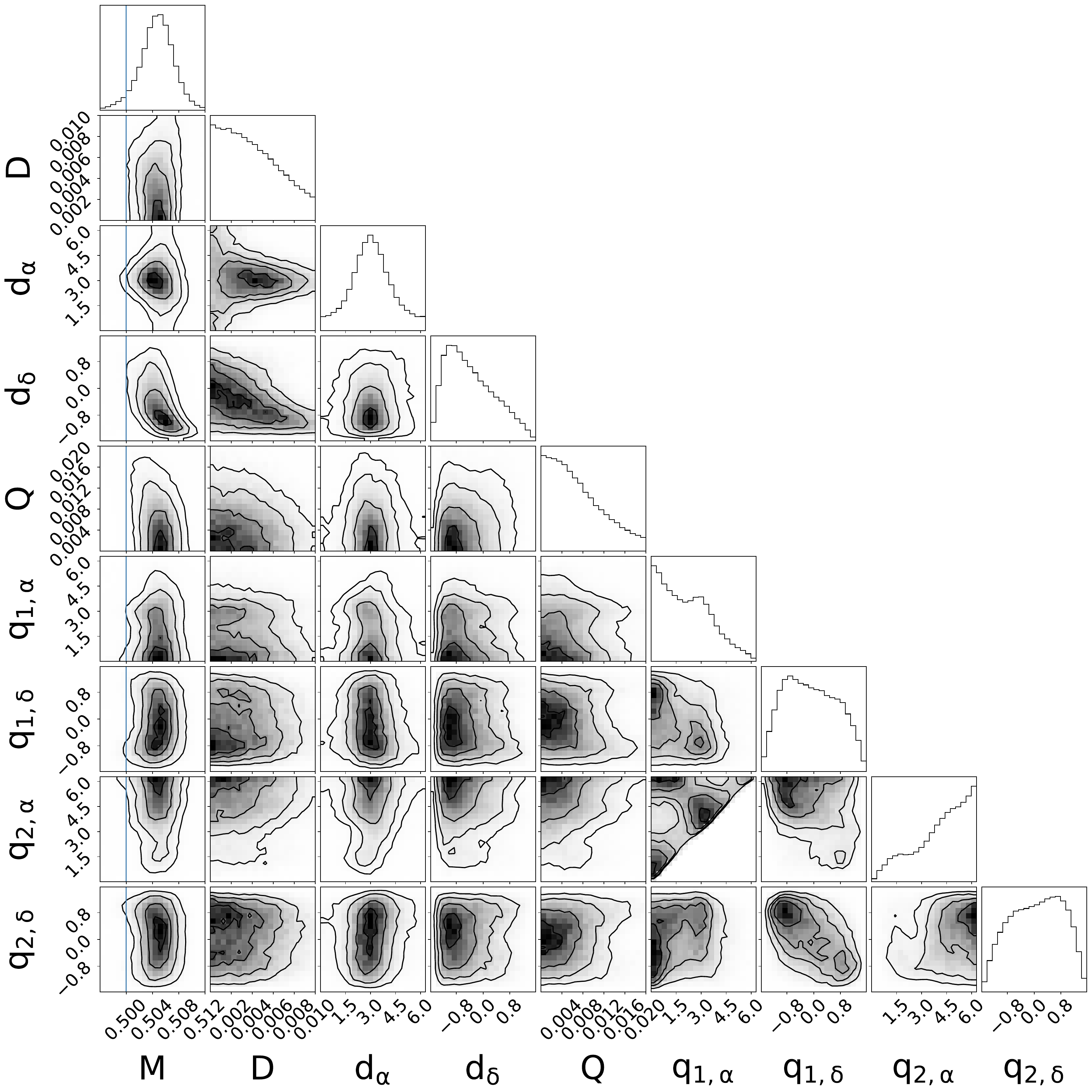}
  \caption{\emph{Left panel:} The monopole plus dipole inference on \texttt{GAN M}. \emph{Right panel:} Also inferring the quadrupole. The blue vertical lines indicate $M=0.5$.}
  \label{fig:corner}
\end{figure*}

\begin{table}
\centering
\caption{Table of parameter constraints when inferring $M$ alone. Limits are at $1\sigma$.}
\label{tab:M}
\renewcommand{\arraystretch}{1.5} 
\begin{tabular}{|l|c|c|}
\hline
\textbf{Dataset} & $\mathbf{M}$ \\
\hline
\texttt{Longo} & $0.512^{+0.004}_{-0.004}$ \\
\texttt{Iye} & $0.503^{+0.002}_{-0.002}$ \\
\texttt{SDSS DR7} & $0.501^{+0.006}_{-0.006}$ \\
\texttt{GAN M} & $0.505^{+0.001}_{-0.001}$ \\
\texttt{GAN NM} & $0.497^{+0.001}_{-0.001}$ \\
\texttt{Shamir} & $0.503^{+0.002}_{-0.002}$ \\
\texttt{PS DR1} & $0.510^{+0.003}_{-0.003}$ \\
\hline
\end{tabular}
\end{table}

\begin{table}
\centering
\caption{Table of parameter constraints when inferring $D$ alone (with $M$ fixed to 0.5). $\Delta$BIC is relative to the monopole-only model; the positive values indicate that the inclusion of dipole parameters is not warranted by the data.}
\label{tab:D}
\renewcommand{\arraystretch}{1.5} 
\begin{tabular}{|l|c|c|c|}
\hline
\textbf{Dataset} & $\mathbf{D}$ & \textbf{$\Delta$BIC} \\
\hline
\texttt{Longo} & $0.020^{+0.006}_{-0.006}$ & 14.5 \\
\texttt{Iye} & $<$ 0.006 & 21.7 \\
\texttt{SDSS DR7} & $<$ 0.019 & 17.5 \\
\texttt{GAN M} & $<$ 0.008 & 39.6 \\
\texttt{GAN NM} & $<$ 0.005 & 24.4 \\
\texttt{Shamir} & $<$ 0.007 & 26.2 \\
\texttt{PS DR1} & $0.020^{+0.006}_{-0.006}$ & 31.7 \\
\hline
\end{tabular}
\end{table}

\begin{table}
\centering
\caption{Results inferring $M$ and $D$ simultaneously.}
\label{tab:long}
\renewcommand{\arraystretch}{1.5} 
\begin{tabular}{|l|c|c|c|c|c|}
\hline
\textbf{Dataset} & $\mathbf{M}$ & $\mathbf{D}$ & \textbf{$\Delta$BIC} & \textbf{\textit{p}-value} \\
\hline
\texttt{Longo} & $0.500^{+0.027}_{-0.027}$ & $<$ 0.016 & 28.9 & 0.44 \\
\texttt{Iye} & $0.503^{+0.002}_{-0.002}$ & $<$ 0.005 & 33.6 & 0.92 \\
\texttt{SDSS DR7} & $0.501^{+0.025}_{-0.028}$ & $<$ 0.046 & 24.7 & 0.32 \\
\texttt{GAN M} & $0.505^{+0.002}_{-0.002}$ & $<$ 0.006 & 35.0 & 0.29 \\
\texttt{GAN NM} & $0.497^{+0.002}_{-0.002}$ & $<$ 0.004 & 35.5 & 0.91 \\
\texttt{Shamir} & $0.503^{+0.002}_{-0.002}$ & $<$ 0.006 & 30.6 & 0.84 \\
\texttt{PS DR1} & $0.509^{+0.003}_{-0.003}$ & $0.016^{+0.006}_{-0.007}$ & 20.2 & 0.054 \\
\hline
\end{tabular}
\end{table}

\begin{table}
\centering
\caption{Results inferring $M$, $D$ and $Q$ simultaneously.}
\label{tab:long2}
\renewcommand{\arraystretch}{1.5} 
\begin{tabular}{|l|c|c|c|c|c|}
\hline
\textbf{Dataset} & $\mathbf{M}$ & $\mathbf{D}$ & $\mathbf{Q}$ & \textbf{$\Delta$BIC} \\
\hline
\texttt{Longo} & $0.499^{+0.009}_{-0.012}$ & $<$ 0.023 & $0.070^{+0.025}_{-0.026}$ & 76.4 \\
\texttt{Iye} & $0.504^{+0.002}_{-0.002}$ & $<$ 0.005 & < 0.009 & 87.3 \\
\texttt{SDSS DR7} & $0.500^{+0.038}_{-0.039}$ & $<$ 0.066 & $<$ 0.090 & 67.2 \\
\texttt{GAN M} & $0.505^{+0.002}_{-0.003}$ & $<$ 0.006 & $<$ 0.010 & 94.9 \\
\texttt{GAN NM} & $0.497^{+0.002}_{-0.002}$ & $<$ 0.004 & $<$ 0.011 & 93.7 \\
\texttt{Shamir} & $0.503^{+0.002}_{-0.002}$ & $<$ 0.006 & $<$ 0.011 & 87.3 \\
\texttt{PS DR1} & $0.510^{+0.004}_{-0.004}$ & $0.017^{+0.007}_{-0.007}$ & $<$ 0.021 & 77.9 \\
\hline
\end{tabular}
\end{table}

\begin{table}
\centering
\caption{Hemispherical anisotropy results, inferring $M$ and $A$ simultaneously.}
\label{tab:long_5}
\renewcommand{\arraystretch}{1.5} 
\begin{tabular}{|l|c|c|c|c|}
\hline
\textbf{Dataset} & $\mathbf{M}$ & $\mathbf{A}$ & \textbf{$\Delta$BIC} \\
\hline
\texttt{Longo} & $0.494^{+0.006}_{-0.005}$ & $0.013^{+0.006}_{-0.006}$ & 28.9 \\
\texttt{Iye} & $0.503^{+0.002}_{-0.002}$ & $<$ 0.004 & 33.6 \\
\texttt{SDSS DR7} & $0.502^{+0.124}_{-0.128}$ & $<$ 0.125 & 24.6 \\
\texttt{GAN M} & $0.505^{+0.002}_{-0.002}$ & $<$ 0.003 & 35.5 \\
\texttt{GAN NM} & $0.497^{+0.002}_{-0.002}$ & $<$ 0.002 & 34.2 \\
\texttt{Shamir} & $0.503^{+0.002}_{-0.002}$ & $<$ 0.004 & 33.2 \\
\texttt{PS DR1} & $0.510^{+0.003}_{-0.003}$ & $0.009^{+0.003}_{-0.004}$ & 30.7 \\
\hline
\end{tabular}
\end{table}


Our results are presented in Tables~\ref{tab:M}--\ref{tab:long_5}. We see that $M$ is consistent with 0.5 within $\sim$3$\sigma$ regardless of whether or not one infers $D$ or $Q$, indicating no significant direction-independent bias in the assignment of Z-wise versus S-wise spins. (To more decimal places, the \texttt{GAN M} result is $0.50532\pm0.00135$, a 3.9$\sigma$ difference from 0.5.) Such biases in annotation methods are well documented, for example for visual assessment by citizen scientists in~\citet{Land, 2009MNRAS.392.1225S, 2017MNRAS.466.3928H}.
This may be at play to a minor degree in the \texttt{Longo}, \texttt{GAN M} and \texttt{PS DR1} datasets.

When inferring $D$ alone, we see a detection of a dipole at just over $3\sigma$ in the \texttt{Longo} and \texttt{PS DR1} datasets, with direction $d_{\alpha}=3.62^{+0.48}_{-0.52}, d_{\delta}=0.51^{+0.46}_{-0.46}$ and $d_{\alpha}=4.57^{+0.30}_{-0.38}, d_{\delta}=0.74^{+0.29}_{-0.31}$ respectively. The remainder have $D$ consistent with 0 at $2\sigma$, such that we present only upper limits (and hence there are no meaningful constraints on the dipole direction). These constraints are fairly tight, indicating that a sizeable dipole can be ruled out at high confidence. The positive $\Delta$BIC for all datasets relative to the monopole-only case indicates a worse-fitting model. The significance of $\Delta$BIC can be interpreted on the Jeffreys scale~\citep{Jeffreys}, which rates the evidence for the better-fitting model as ``decisive'' if the Bayes factor (ratio of evidences) exceeds 100. Since evidence $\approx\exp(-\text{BIC}/2)$, this corresponds to $\Delta$BIC>9.2. Thus in all cases the evidence in favour of the monopole-only model is decisive.

From Table~\ref{tab:long} we see that it is no coincidence that \texttt{Longo} and \texttt{PS DR1} have separate monopole and dipole detections: when inferring both $M$ and $D$, both \texttt{Longo} anomalies disappear, while those of \texttt{PS DR1} are reduced in significance, the dipole to almost 2$\sigma$. This illustrates the argument of~\citet{Land} that degeneracies between $M$ and $D$ require them to be inferred jointly. The small remaining \texttt{PS DR1} dipole points towards $d_{\alpha}=4.03^{+0.28}_{-0.34}, d_{\delta}=0.28^{+0.42}_{-0.44}$. Even for \texttt{PS DR1} the $\Delta$BIC of the monopole+dipole model is $>0$, indicating that the inclusion of a dipole is not warranted by the Bayesian evidence. The amount by which any additional parameter must increase the likelihood to be warranted is fairly high due to the large sizes of the datasets.

Moving onto Table~\ref{tab:long2}, we see that only \texttt{Longo} has a non-zero quadrupole at $2\sigma$, with direction $q_{1, \alpha}=1.63^{+0.78}_{-0.78}, q_{1, \delta} = 0.15^{+0.46}_{-0.48}, q_{2, \alpha} = 4.72^{+0.75}_{-0.81}, q_{2, \delta}=-0.14^{+0.48}_{-0.47}$. This is however not significant at the $3\sigma$ level. For the other samples the $Q$ bounds are tight, leading to no significant inflation of the $D$ bounds. The $\Delta$BIC values are even larger than for the monopole+dipole model due to the inclusion of a further five unwarranted parameters. We show the full corner plots for \texttt{GAN M}, which had the highest claimed dipole significance (Table~\ref{Table 1}), for the $M$+$D$ and $M$+$D$+$Q$ analyses in Fig.~\ref{fig:corner}.

The preference for $q_{1, \alpha}$ near 0 and $q_{2, \alpha}$ near $2\pi$ is a volume effect due to the flat priors, which occurs because $Q\approx0$ so the quadrupole angles are not constrained by the likelihood. Having $q_{1, \alpha}$ near 0 allows $q_{2, \alpha}$ to take any value in [0,$2\pi$), while conversely have $q_{2, \alpha}$ near $2\pi$ allows $q_{1, \alpha}$ to take any value in [0,$2\pi$), due to the requirement that they are ordered. This also explains why the $q_{1, \alpha}$ and $q_{2, \alpha}$ posteriors are not symmetric as RA wraps around $2\pi$ back to 0: when this happens, $q_{1, \alpha}$ and $q_{2, \alpha}$ interchange. The priors also account for the fact that the sums of the average $q_{1, \alpha}$, $q_{2, \alpha}$ values are typically near $2\pi$, and $q_{1, \delta}$, $q_{2, \delta}$ near 0. When $Q$ is constrained to be near 0, the axis directions follow their priors which are centred on $\pi$ for RA and 0 for Dec. However, this is also the case for the \texttt{Longo} dataset, where there is a $>2\sigma$ detection of $Q$. This indicates that the two quadrupole axes are roughly anti-aligned, creating angular variation like a dipole with a cosine-squared as opposed to cosine angular variation.

To investigate any potential redshift-dependence of the results, we repeat the inference of $M$ and $D$ for \texttt{Longo} (the only one that provides redshift information) separately for galaxies in the ranges $0.000064<z<0.04183$, $0.04183<z<0.063036$ and $0.063045<z<0.084998$. This puts an equal number of galaxies into each tomographic bin. In all cases we find constraints consistent with the full \texttt{Longo} dataset (and the others) that $M\approx0.5$ and $D\approx0$, and that the posteriors are very similar to the results with the spins randomised.

Finally, in Table~\ref{tab:long_5} we see that in all datasets except two $A$ is consistent with 0 within $2\sigma$, and hence present only upper limits. The exceptions (at very slightly over $2\sigma$) are \texttt{PS DR1}, which has a minor preference for hemispherical anisotropy in the direction $a_{\alpha}=4.99^{+0.19}_{-0.96}, a_{\delta} = -0.038^{+0.69}_{-0.33}$, and \texttt{Longo}, which has a preference in the direction $a_{\alpha}=0.92^{+0.76}_{-0.16}, a_{\delta} = -0.050^{+0.29}_{-0.51}$. However, the $\Delta$BIC values of the anisotropic relative to isotropic models are similar to those in Table~\ref{tab:long}, indicating that this form of asymmetry is not warranted by the Bayesian evidence either.

\subsection{Frequentist analysis}
\label{sec:res_freq}

\begin{figure}
  \centering
  \includegraphics[width=0.50\textwidth]{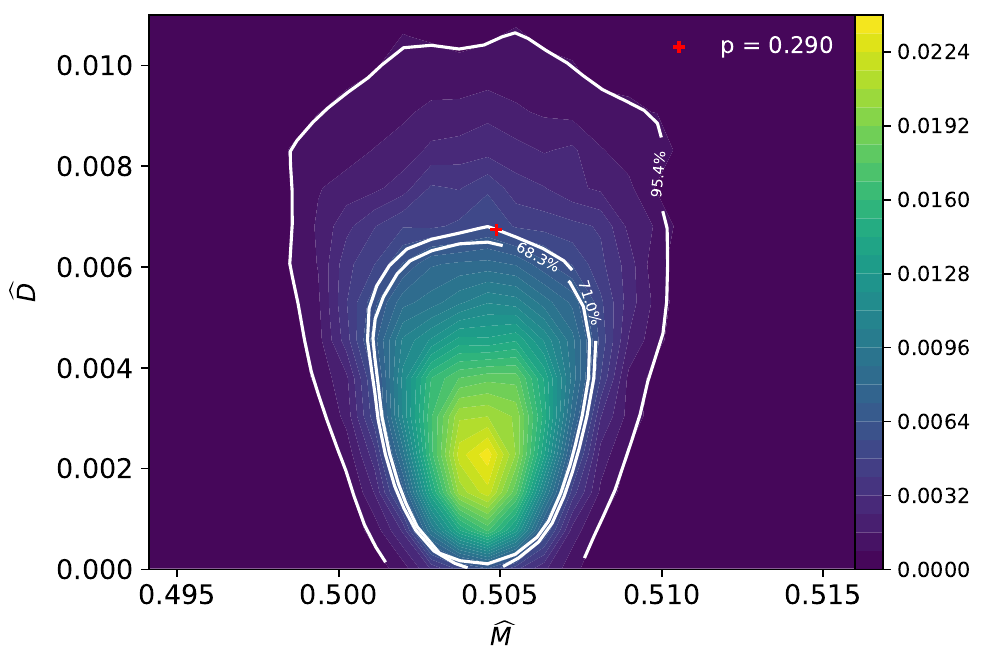}
  \caption{The heatmap shows the distribution of maximum-likelihood $M$ and $D$ values for 50,000 mock datasets with the same galaxy positions as in \texttt{GAN M}, but with spins selected from the model with $M=0.505$, $D=Q=0$. The maximum-likelihood value in the real data is shown by the red plus symbol. 26.6 per cent of the isotropic mock datasets are more extreme than this.}
  \label{fig:GAN_M_freq}
\end{figure}

\begin{figure}
  \centering
  \includegraphics[width=0.50\textwidth]{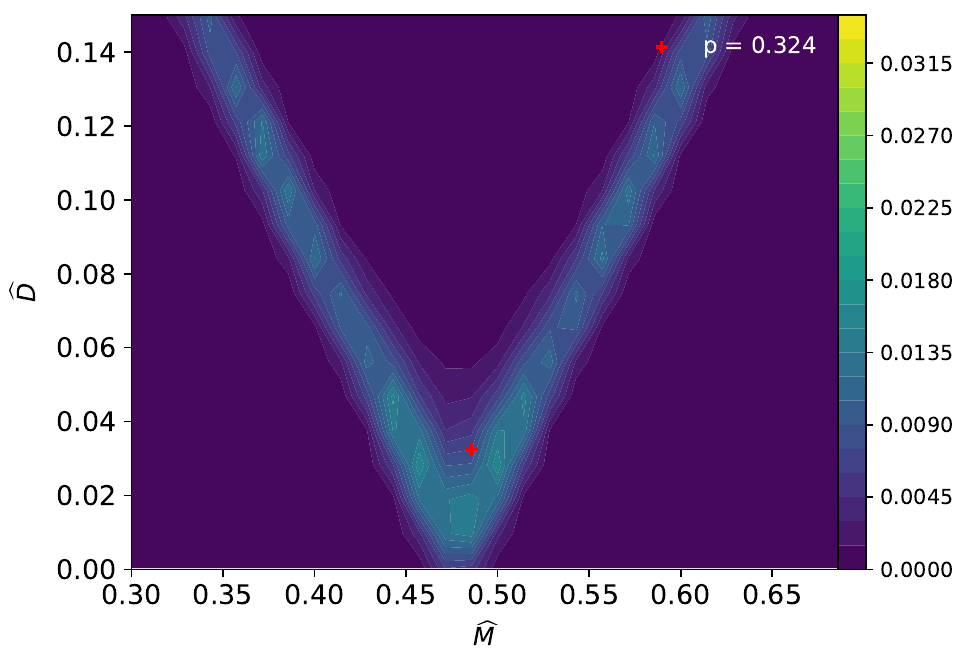}
  \caption{As Fig.~\ref{fig:GAN_M_freq}, but for \texttt{SDSS DR7}. In this case the mock data are generated according to the maximum-likelihood value of $M$, 0.486, and $D=Q=0$. The contour lines are suppressed for clarity.}
  \label{fig:SDSS_freq}
\end{figure}

In the final column of Table~\ref{tab:long} we show the $p$-value of the null hypothesis of isotropy, calculated using mock data generated according to $M=\widehat{M}$, $D=Q=0$ (see Sec.~\ref{sec:meth_freq}). $\texttt{PS DR1}$ nears the 2$\sigma$ level of $p=0.05$, although the BIC still clearly indicates that the monopole-only model is preferred. The frequentist analysis therefore corroborates the Bayesian one that there is no significant evidence for anisotropy. The method is illustrated for \texttt{GAN M} in Fig.~\ref{fig:GAN_M_freq}, in which the distribution of recovered $M$ and $D$ values on the mock datasets are compared to those of the real data.

It is worth emphasising that the patchy sky coverage of some of our datasets lead to significant parameter degeneracies, which both of our analysis methods naturally account for. In particular, \texttt{SDSS DR7} consists of a relatively small number of galaxies with poor sky coverage (see Fig.~\ref{fig:skyplot}), and hence cannot distinguish between a modified monopole and a dipole aligned or antialigned with the observed region. To illustrate the effect of this, we show in Fig.~\ref{fig:SDSS_freq} the counterpart of Fig.~\ref{fig:GAN_M_freq} for this dataset. Our analysis would correctly recover no significant anisotropy even if the best-fit $M$ and $D$ values in real data were far from $0.5$ and $0$ in the degeneracy direction.

\subsection{Comparison to the literature}
\label{sec:res_lit_comp}

Our clear findings in support of galaxy spin isotropy raise the question of why others have reached diametrically opposite conclusions. To investigate this, we attempt to implement the methods of some such authors on their respective datasets.

The only available mention of a dipole statistic in studies claiming a dipole is eq. 1 of \citet{McAdam_Shamir_2023_GAN}, which in our notation reads
\begin{equation}
\label{eq:shamir_chi2}
\chi^2_{\alpha,\delta} = \sum_i \left|\frac{\left(s_i |\vec{d}_{\alpha,\delta} \cdot \vec{n}_i| - \vec{d}_{\alpha,\delta} \cdot \vec{n}_i\right)^2}{\vec{d}_{\alpha,\delta} \cdot \vec{n}_i}\right|.
\end{equation}
$\vec{d}_{\alpha,\delta}$ is the unit dipole axis in the $\alpha,\delta$ direction. This $\chi^2$ is evaluated on a grid of $\alpha, \delta$ for the real data (yielding $\chi^2_{\alpha,\delta,\text{data}}$) and also of 1000 mock data sets in which the spin directions are randomised (yielding $\chi^2_{\alpha,\delta,\text{mock},i}$ for the $i^\text{th}$ mock data set). The significance of the dipole in the direction of $\alpha, \delta$ is then calculated as
\begin{equation}
\sigma_{\alpha,\delta} = \frac{|\chi^2_{\alpha,\delta,\text{data}} - \langle{\chi^2_{\alpha,\delta,\text{mock},i}}\rangle|}{\text{std}(\chi^2_{\alpha,\delta,\text{mock},i})}
\end{equation}
where angled brackets denote a mean over the mock data sets. Eq.~\ref{eq:shamir_chi2} appears to be Pearson's $\chi^2$ statistic, in which one replaces the squared uncertainty in the denominator of the regular Gaussian $\chi^2$ by the expected value, in this case (something proportional to) $\vec{d} \cdot \vec{n}_i$.
However, the observed value is $s_i$, not $s_i |\vec{d} \cdot \vec{n}_i|$ which mixes the observation with the expectation. This effectively projects $s_i$ onto the dipole axis, which amounts to modelling the expected value as 1 everywhere in the hemisphere aligned with the dipole direction, neglecting the fact that the likelihood of $s=1$ is lower the further one is from the dipole axis, even if the expected value is $>0.5$. Larger discrepancies from the dipole axis may contribute more to the overall chi-squared statistic. Even besides this, we do not consider Eq.~\ref{eq:shamir_chi2} a useful statistic because it does not capture the sampling distribution of the observable as do both our Bayesian and frequentist methods. Furthermore, our attempt at using this equation on the \citet{McAdam_Shamir_2023_GAN} dataset did not yield the results quoted in that paper, so we were unable to reproduce their analysis. An attempt to reproduce the results of \citet{Longo} using Eq.~\ref{eq:shamir_chi2} (a shot in the dark, since \citealt{Longo} do not define their $\chi^2$ statistic) similarly failed.

\section{Conclusion}
\label{sec:conc}

We have analysed seven datasets of galaxy sky positions and spin directions to assess the evidence for anisotropy in galaxies' angular momenta. Four of these datasets have literature claims of a $>$2$\sigma$ dipole in the spin directions, with two at $>$3$\sigma$. However, we find clear consistency with statistical isotropy in all datasets using either a Bayesian or frequentist method, both of which account for the look-elsewhere effect and account fully for parameter degeneracies. Due to the incomplete sky coverage spherical harmonics are not orthogonal, leading us to explore the possibility of a quadrupole as well as a dipole and monopole, but this too is small and does not affect our $\ell=0$ or $\ell=1$ results. The evidence for anisotropy does not increase if the cosine dependence of the dipole is removed. We trace the difference with literature results claiming a dipole to the unmotivated statistics that they employ, and do not find their results to be reproducible.

In conclusion, galaxy spins exhibit large-scale isotropy in adherence to the cosmological principle. Our work highlights the vital importance of careful statistics in analysing fundamental properties of the Universe.

\section*{Data availability}

The annotated galaxy catalogues are available online at the following URLs:
\begin{itemize}
\item \texttt{Longo}: \url{https://ars.els-cdn.com/content/image/1-s2.0-S0370269311003947-mmc1.txt}
\item \texttt{Iye}: \url{https://people.cs.ksu.edu/~lshamir/data/iye_et_al/}
\item \texttt{SDSS DR7}: \url{https://people.cs.ksu.edu/~lshamir/data/sdss_phot/}
\item \texttt{GAN M/NM}: \url{https://people.cs.ksu.edu/~lshamir/data/SpArcFiRe/}
\item \texttt{Shamir}: \url{https://people.cs.ksu.edu/~lshamir/data/assymdup/}
\item \texttt{PS DR1}: \url{https://people.cs.ksu.edu/~lshamir/data/assym3/}; \url{https://figshare.com/articles/dataset/PanSTARRS_DR1_Broad_Morphology_Catalog/12081144}
\end{itemize}
Our code is available on github \githublink. Any other data underlying the article will be made available on reasonable request to the authors.

\section*{Acknowledgements}

We thank Pedro Ferreira, Kazuya Koyama and Sebastian von Hausegger for useful discussions.

DP was supported by a SEPnet Summer Placement at the Institute of Cosmology and Gravitation, University of Portsmouth. HD is supported by a Royal Society University Research Fellowship (grant no. 211046).

This project has received funding from the European Research Council (ERC) under the European Union's Horizon 2020 research and innovation programme (grant agreement No 693024). For the purpose of open access, we have applied a Creative Commons Attribution (CC BY) licence to any Author Accepted Manuscript version arising.

\bibliographystyle{mnras}
\bibliography{references}


\label{lastpage}
\end{document}